\documentclass[RNAAS]{aastex631}
\usepackage{CJK}

\usepackage{verbatim, amsmath, amsfonts, amssymb,amssymb}
\usepackage{multirow}
\usepackage[utf8]{inputenc}
\usepackage{empheq}
\usepackage[skins,theorems]{tcolorbox}
\usepackage{latexsym}
\usepackage{bigints}
\usepackage{natbib}
\usepackage{color}
\usepackage{tikz}

\shorttitle{SVD denoising}
\shortauthors{Chen \& de Souza }
\begin{document}
 \begin{CJK*}{UTF8}{gbsn}
\title{\texttt{yonder}: A python package for data denoising and reconstruction}

\correspondingauthor{Rafael S. de Souza}
\email{drsouza@shao.ac.cn}

\author[0000-0002-9925-4371]{Peng Chen (彭琛)}
\affiliation{Shanghai Institute of Technology, 100 Haiquan Rd., Shanghai 201418, China}

\author[0000-0001-7207-4584]{Rafael S. de Souza}
\affiliation{Key Laboratory for Research in Galaxies and Cosmology, Shanghai Astronomical Observatory,
Chinese Academy of Sciences, 80 Nandan Rd., Shanghai 200030, China}

\begin{abstract}
We present a standalone implementation of a data-deconvolution method based on singular value decomposition. The tool is written in python and packaged in the open-source  \texttt{yonder} package. \texttt{yonder} receives as input two matrices, one for the data and another for the errors,  and outputs a denoised version of the original dataset. In this Research Note, we briefly describe the methodology and show a demonstration of the \texttt{yonder} on a simulated dataset. 
\end{abstract}


\section{Introduction}

Measurement errors are ubiquitous in Astronomy. While plenty of methodologies exist to handle them on linear regression problems \citep{Kelly2007}, there is a shortage of ready-to-use packages to perform data pre-processing and machine learning-related tasks in the presence of measurement errors \citep[but see, e.g.][]{Bovy2011, Reis_2018}. It is of particular interest to estimate a latent matrix $X$ based on noisy observations. Here, we consider the case of tabular-like data representing different objects as rows and their particular features as columns, and a matrix of similar dimensions encodes the uncertainties of each measurement.  $\mathcal{X}_{obs} = \mathcal{X} + \mathcal{X}_{sd}$, where $\mathcal{X}_{obs}$ is the noisy measurement, and $\mathcal{X}_{sd}$ a stochastic noise matrix with zero mean. To help alleviate this problem, we wrote  \texttt{yonder},  a python package for data denoising and reconstruction. This report briefly introduces the package and illustrates how the method works in practice.

\section{Analysis and Discussion}
\label{sec:svd}
\begin{figure}
\includegraphics[width=0.9\linewidth]{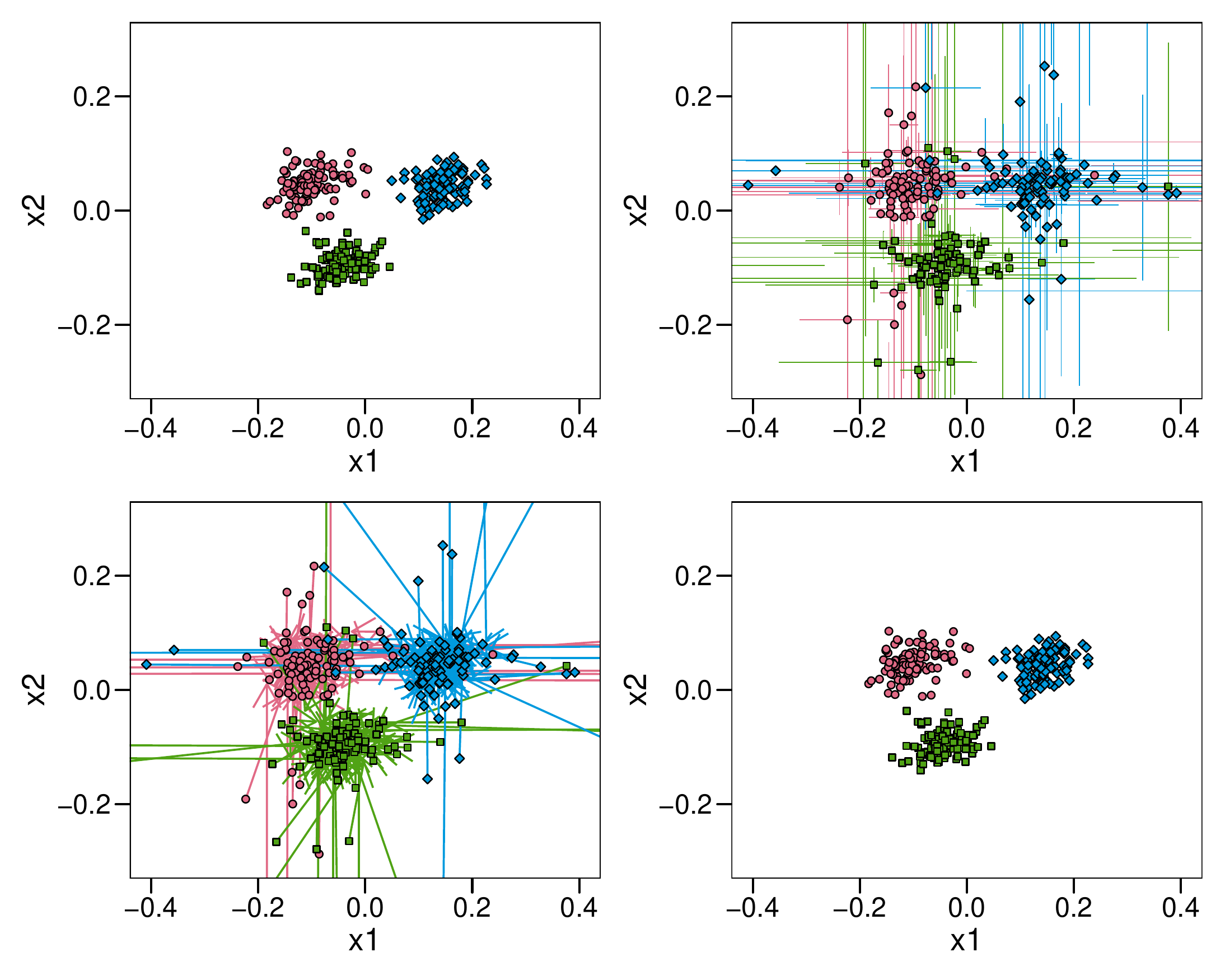}
\caption{Example of the effects of measurement errors on mock data. Top left: scatter plot of the error-free data;  top right: scatter plot with the noisy mock data and respective error-bars; bottom left: scatter plot of the noisy mock data, with arrows pointing from the data to the predicted position by yonder; bottom right: scatter plot of the denoised data.}
\label{fig:yonder}
\end{figure}

Given a  data matrix $\mathcal{X}$,  its singular value decomposition (SVD) is given by:
\begin{equation}
\mathcal{X} =  \mathcal{U}\Sigma \mathcal{V}^{\intercal}, 
\end{equation}
Where $\mathcal{U}\Sigma$ gives the principal components, and the columns of $\mathcal{V}$  the corresponding coefficients of the linear combination of the original variables. In the case of a nosy matrix $\mathcal{X}_{obs}$ with respective measurement errors $\mathcal{X}_{sd}$, there is an iterative solution described in \citet{WENTZELL1999, Wentzell2012}. \texttt{yonder} implements the case of independent and identically distributed heteroscedastic errors.  To illustrate how \texttt{yonder} works, we simulate a noise model similar to \citet{Wentzell2012}, with errors-in-measurements sampled from a log-normal distribution. 
\autoref{fig:yonder} illustrates the workflow. The mock data $\mathcal{X}_{obs_{nm}} = \mathcal{X}_{nm} + \mathcal{N}(0,\sigma_{nm}^2)$ consists of m = 20 columns,  n = 300 rows, and three distinct groups with 100 members each.  The top left panel shows the error-free data: the top right panel the noisy data and respective error bars. \texttt{yonder} uses as input the noisy matrix and the associated error matrix and approximates the position of the unobserved denoised data. The presence of errors blurs the discrimination of the groups, which is recovered after running \texttt{yonder}. The method will be more beneficial for cases of highly correlated data for which a low-rank approximation is guaranteed, for example, in multi-band photometric surveys.  

The code takes an input matrix $\mathcal{X}_{obs}$, and respective error matrix $\mathcal{X}_{sd} \equiv \sigma_{nm}$, and outputs the SVD of the latent denoised matrix  $\hat{\mathcal{X}} = \hat{\mathcal{U}}\hat{\Sigma} \hat{\mathcal{V}}^{\intercal} \approx \mathcal{X}$.

This work showcases that accounting for uncertainties can further benefit machine learning algorithms by correcting the data projection and enhancing the distinguishability between groups. Possible applications include the use in multi-band surveys to find stellar clusters in color-color and colour-magnitude diagrams \citep{Chies-Santos2022}. 

\subsection{Installation and similar software}

\texttt{yonder} is available as an open-source software package on  GitHub\footnote{\href{https://github.com/pengchzn/yonder}{{https://github.com/pengchzn/yonder}}}  and Zenodo\footnote{\href{https://doi.org/10.5281/zenodo.6321520}{https://doi.org/10.5281/zenodo.6321520}}. The code can be installed via pip command\footnote{\url{https://pypi.org/project/yonder/}}.  
Additionally to this \texttt{python} version, there is also a \texttt{R} implementation of the same algorithm available on CRAN\footnote{\href{https://CRAN.R-project.org/package=RMLPCA}{https://CRAN.R-project.org/package=RMLPCA}}.

\software{scipy \citep{Virtanen2020}, numpy \citep{harris2020array}.}

\bibliography{ref}{}
\bibliographystyle{aasjournal}

\end{CJK*}
\end{document}